\documentclass[11pt,cleanfoot,preprint]{asme2ej}
\usepackage{xr}
\usepackage{epsfig}
\usepackage{graphicx}
\usepackage{fancyhdr}
\usepackage{setspace}
\usepackage{helvet}
\usepackage{xcolor}
\usepackage[hyphens]{url}
\usepackage{bm}
\usepackage{amsmath}
\usepackage{mathtools}
\usepackage{graphicx}
\usepackage{dcolumn}
\usepackage[colorlinks=true,linkcolor=blue]{hyperref}%
\makeatletter
\renewcommand\@dotsep{400}
\makeatother

\author{Subhasisa Rath\thanks{Corresponding Author: \href{mailto:subhasisa.rath@gmail.com}{subhasisa.rath@gmail.com} ORCID:\href{https://orcid.org/0000-0002-4202-7434}{0000-0002-4202-7434} [Address all correspondence related to ASME style format and figures to this author]} 
	\affiliation{
		School of Energy Science \& Engineering \\Indian Institute of Technology Kharagpur \\Kharagpur 721302, West Bengal, India
	}	
}

\author{Bimalendu Mahapatra\thanks{Corresponding Author: \href{mailto:bimalendumahapatra1994@gmail.com}{bimalendumahapatra1994@gmail.com} ORCID:\href{https://orcid.org/0000-0003-0845-9775}{0000-0003-0845-9775} [Address all correspondence for other issues to this author]}
	\affiliation{
		Department of Mechanical Engineering \\Indian Institute of Technology Kharagpur\\ Kharagpur 721302, West Bengal, India
	}	
}

\title{Low Reynolds number pulsatile flow of a viscoelastic fluid through a channel: Effects of fluid rheology and pulsation parameters}
\begin{document}
\maketitle    
\doublespacing
\begin{abstract}
{\noindent \it As the first endeavour, we have analyzed the pulsatile flow of Oldroyd-B viscoelastic fluid where the combined effects of fluid elasticity and pulsation parameters on the flow characteristics are numerically studied at a low Reynolds number. Computations are performed using a finite-volume based open-source solver OpenFOAM\textsuperscript{\textregistered} by appending the log-conformation tensor approach to stabilize the numerical solution at high Deborah number. Significant flow velocity enhancement is achieved by increasing the viscoelastic behaviour of the fluid. High-velocity gradient zones and high polymeric stress regions are observed near the channel wall. The magnitude of axial velocity attenuates with increasing pulsation amplitude or pulsation frequency, and the extent of this attenuation is highly dependent on the Deborah number or the retardation ratio. This work finds application in the transport of polymeric solutions, extrusion, and injection moulding of polymer melts in several process industries.}
\end{abstract}

\section{Introduction}
{Pulsatile flow of viscoelastic fluids plays an indispensable role in various process industries as well as in biomedical and biochemical applications \cite{Jia2008,shah1996pulsating,wang2007analysis,phan1978pulsating,sankar2009mathematical}. It promotes inherent advantages in mixing and separations of chemical species, transport of polymeric solutions or suspensions, injection moulding of polymer melts, blood flow through arteries and veins, and drug delivery \cite{phan1982pulsating,jinping2008influence,razavi2011numerical,tu1996pulsatile,hettiarachchi2011effect,dhananchezhiyan2018improving,mahapatra2020electroosmosis}. Viscoelastic fluids are the class of non-Newtonian fluid where some polymers are blend with a solvent. Thus, viscoelastic fluids exhibit both viscous and elastic behaviours, i.e., elastic recovery after deformation and stress relaxation \cite{Bird1987}. Owing to the time-dependent behaviour or relaxation phenomena and memory effects, the underlying physics of viscoelastic fluids is quite complex. The interplay of complex rheology of the viscoelastic fluid and the pulsatile nature of the flow further accentuate the complexity of the hydrodynamic problem. Hence, proper understanding and characterization of viscoelastic fluids are vital for their applicability in real-life practical scenarios.}
	
{Over the last few decades, a noticeable development in research have been seen for non-Newtonian viscoelastic fluids. Several linear, non-linear, and quasi-linear viscoelastic fluid models have been developed to apprehend the intricate flow physics associated with the elastic behaviour of the complex fluids. For instance, the Giesekus model\cite{Giesekus}, Maxwell model\cite{Maxwell}, Jeffreys model\cite{Jeffreys}, Oldroyd-B model\cite{Oldroyd}, Phan–Thien–Tanner (PTT) model\cite{PTT2}, Finitely Extensible Nonlinear Elastic (FENE) model\cite{FENE}, and Pom–Pom model\cite{PomPom} are few well-known viscoelastic fluid models in literature. In numerical computations, an elementary problem associated with the viscoelastic fluid model is the instability or convergence issue at high Weisenberg number ($Wi$), or Deborah number ($De$)\cite{amoreira}. Hence, a verity of approaches have been proposed to overcome such problem and stabilize the numerical solution such as: log-conformation tensor approach \cite{fattal2004,fattal2005,afonso2009}, both-sides-diffusion technique\cite{Chen,mahapatra2021numerical,mahapatra2021microconfined}, elastic-viscous stress splitting methods\cite{perera,sun}, and solvent-polymer decomposition method\cite{Bird1987,duarte2008}. 
		
{Aiming to better understand the complex flow dynamics of pulsatile flow of biological fluids, several studies were conducted using Newtonian and viscoelastic fluid models \cite{Rosenfeld1995,Daidzic2014,Budwig1997,Yapici2012,Jia2008}. Cable and Boger \cite{cable1979} experimentally investigated the unstable abrupt entry flow of viscoelastic fluids in a tube using a periodic spiralling frequency method. The periodic entry flow disturbances, criteria for the onset of periodic disturbances, and random entry flow disturbances were elaborately explained in their study. The oscillating and pulsating flow of viscoelastic fluids were experimentally studied by Khabakhpasheva $et~al.$ \cite{khabakhpasheva1989} using a sinusoidal time-varying pressure gradient. Axial velocity profiles as a function of the pulsation phase are plotted and compared with the pressure drop profiles. An enhancement in flow rate has been observed compared to a steady flow, and the flow enhancement with amplitude of pulsation for different ranges of $Re$, $De$, and $Wi$ were discussed in their study. Letelier $et~al.$ \cite{letelier2002} and Siginer and Letelier \cite{siginer2002} analytically solved the pulsatile flow of viscoelastic fluids in different cross-section of straight tubes to delineate the longitudinal field and secondary flows, respectively. Duarte $et~al.$ \cite{duarte2008} numerically and analytically investigated both start-up and pulsating test case problem for unsteady viscoelastic flows. A periodic pressure gradient has been superimposed on the constant Poiseuille flow to demonstrate the pulsatile flow problem using various viscoelastic fluid models. Rachid and Ouazzani \cite{rachid2014} analytically investigated the effect of pulsatile flow on low $Re$ peristaltic transport of a viscoelastic fluid in a cylindrical tube using the Maxwell model.} 
			
{Pulsatile flow of viscoelastic fluids plays a vital role in biochemical and biomedical applications to mimic the blood flow in arteries, veins, and vessels. Javadzadegan \cite{javadzadegan2009} numerically studied the pulsatile flow of both viscous and viscoelastic fluids in a constricted stenosed artery using oscillatory pressure gradient with Oldroyd-B and Cross models. Flow rate, velocity profiles, and wall shear-stress have been clearly elaborated in their study. Bakhti $et~al.$ \cite{bakhti2017} numerically investigated the pulsatile blood flow through a constricted tapered artery using the generalized Oldroyd-B model with variable-order fractional derivative technique. Moyers $et~al.$ \cite{moyers2009} analytically solved the high-frequency oscillatory blood flow in a tube using the non-homogeneous	hemorheological model. The amplitude relationship between the pressure gradient and volume flow rate has been investigated in their study. Sajid $et~al.$ \cite{sajid2015} numerically demonstrated the effect of body acceleration on the pulsatile flow of blood in a vessel using the Oldroyd-B viscoelastic model. To elucidate the effects of stress relaxation and retardation in unsteady pulsatile flow was the primary aim of their work.}
			
{The previous studies on the pulsatile flow of viscoelastic fluids have revealed many physical insights into the flow physics where a time-varying pressure gradient was imposed to generate the pulsating flow. The time-varying pressure gradient gives rise to a time-varying velocity pulsation in the flow, which may or may not be in the same phase depending upon the flow conditions \cite{khabakhpasheva1989}. As the first endeavour in the field of viscoelastic fluids, a sinusoidal axial velocity pulsation has been superimposed on the mean flow to trigger a pulsatile flow in the channel, albeit research has been done in Newtonian and power-law viscoplastic fluids \cite{qamar2011,gupta2020}. In real life practical experiments, one can convert the velocity to flow rate and use the time-varying flow rate to mimic the pulsatile flow in a duct or channel. In the present study, a pulsatile flow of viscoelastic fluid in a parallel plate channel has been studied numerically to elucidate the combined effects of pulsation parameters and complex rheology of the viscoelastic fluid on flow characteristics. This work finds application in the transport of polymeric solutions, extrusion and injection moulding of the polymer melt in several process industries, and transport of biofluids in microfluidic devices.}

\section{Problem description}
{In the present study, we have considered a parallel-plate channel of height $2H$ and length $60H$ to simulate the pulsatile flow of viscoelastic fluid, as shown in Fig. \ref{fig:1}. The origin of the Cartesian coordinate system is located exactly at the centre of the inlet such that the top and bottom walls are placed at $y=\pm H$, respectively. A sinusoidal pulse, $a\sin(\omega t)$, is superimposed on the non-zero mean or base flow $(U_0)$ to generate a time-varying axial pulsatile velocity $U_x$. Where $a$ is the amplitude of pulsation, $\omega$ is the angular frequency of pulsation, and $t$ is the time. The viscoelastic fluid described by the Oldroyd-B constitutive equations flows inside the channel due to the imposition of velocity pulsation. We have illustrated the governing equations, non-dimensionalization of equations, and relevant boundary conditions for the present flow configuration in the forthcoming sections.}


\subsection{Governing equations}
{The dimensional governing equations for conservation of mass and linear-momentum in the absence of any external body force for an incompressible, isothermal, laminar, and two-dimensional flow are given by:}

\begin{equation}\label{eq:1}
	\nabla\cdot\mathbf{U}=0
\end{equation}
\begin{equation}\label{eq:2}
	\rho \frac{\partial \mathbf{U}}{\partial t}+\rho \nabla\cdot(\mathbf{U}\cdot\mathbf{U})=-\nabla p+\nabla\cdot\boldsymbol{\tau}
\end{equation}
Where, $\rho$ denotes the  mass density, $\mathbf{U}$ is the velocity vector having its components $U_x$ and $U_y$ in axial and transverse directions, respectively. The symbol `$p$' represents the fluid pressure and $\boldsymbol{\tau}$ is the stress tensor of the viscoelastic fluid which shows a complex relationship with the strain rate \cite{Bird1987}. In this study, we implement a quasi-linear viscoelastic fluid  more precisely Oldroyd-B model \cite{Oldroyd} to capture the rheological behavior of the fluid. The extra stress tensor $\boldsymbol{\tau}$ for the Oldroyd-B constitutive model is described by:
\begin{equation}\label{eq:3}
	\boldsymbol{\tau}+\lambda \overset{\nabla}{\boldsymbol{\tau}}=2\eta \Big(\mathbf{D}+\lambda(1-\kappa)\overset{\nabla}{\mathbf{D}}\Big)
\end{equation}
In Eq. \ref{eq:3}, $\lambda$ is the fluid relaxation time \cite{vamerzani2014} and $\eta$ is the total viscosity of the fluid, where $\eta=\eta_s+\eta_p$. Here, $\eta_s$ and $\eta_p$ are the solvent and polymeric viscosity contributions at zero shear-rate, respectively. The symbol `$\kappa$' is defined as the ratio of polymeric viscosity to the total viscosity and the fluid retardation time is of the form $\lambda_r=\lambda(1-\kappa)$. The rate of deformation tensor is denoted as $\mathbf{D}$, where $\mathbf{D}$=$\frac{1}{2}[\nabla\mathbf{U}+\nabla\mathbf{U}^\mathrm{T}]$. Finally, $\overset{\nabla}{\boldsymbol{\tau}}$ and $\overset{\nabla}{\mathbf{D}}$ are the upper-convected derivative (UCD) terms of $\boldsymbol{\tau}$ and $\mathbf{D}$. The UCD of any second-order tensor $\mathbf{M}$ is defined as
\begin{equation}\label{eq:4}
	\overset{\nabla}{\mathbf{M}}=\frac{\partial \mathbf{M}}{\partial t}+\mathbf{U}\cdot\nabla\mathbf{M}-\mathbf{M}\cdot\nabla\mathbf{U}-\nabla\mathbf{U}^\mathrm{T}\cdot\mathbf{M}
\end{equation}

\subsection{Non-dimensionalization}
The dimensional governing equations employed to solve the present problem are non-dimensionalized using proper scaling of variables to obtain a dimensionless set of equations and reduced number of variables to define the flow physics. Here, the superscript `$\sim$' denotes the non-dimensional parameters which are scaled using the characteristic values as defined below.
\begin{equation}\label{eq:5}
	\begin{aligned}
		\widetilde{t}=\frac{tU_0}{H}~,~~\widetilde{x}=\frac{x}{H}~,~~\widetilde{y}=\frac{y}{H}~,~~\widetilde{U}_x=\frac{u_x}{U_0}~,~~\widetilde{U}_y=\frac{u_y}{U_0}~,~~\\\widetilde{p}=\frac{pH}{\eta U_0}~,~~\widetilde{\tau}_{xx}=\frac{\tau_{xx}H}{\eta U_0}~,~~\widetilde{\tau}_{xy}=\frac{\tau_{xy}H}{\eta U_0}~,~~\widetilde{\tau}_{yy}=\frac{\tau_{yy}H}{\eta U_0}
	\end{aligned}
\end{equation}
Here, we have considered the advection time to be the characteristic time scale, the half-height of the channel as the characteristic length scale, the mean flow velocity as the characteristic velocity scale and viscous stress to scale the fluid pressure and stress components.
\subsection{Flow governing parameters}
It is worthwhile to define all the governing dimensionless parameters and there range of values associated with the present work. We have defined all the dimensionless parameters and their range of values in Table \ref{tab:parameters}.
Here the Reynolds number is defined as the ratio of inertial forces to the viscous forces based on the mean flow velocity. The ratio of the characteristic time scale, which is the relaxation time ($\lambda$) of the fluid in this study, to the time scale of deformation is termed as the Deborah number \cite{alves2003}. The ratio of the retardation time ($\lambda_r$) to the relaxation time ($\lambda$) of the viscoelastic fluid, which is equal to the ratio of solvent viscosity to the total viscosity of the fluid, is defined as the retardation ratio or viscosity ratio ($\beta$) \cite{duarte2008,alves2003}. The dimensionless pulsatile flow frequency scaled with the viscous effects is known as the Womersley number ($Wo$)\cite{kiran2019}. The dimensionless pulsation amplitude ($A$) is defined as the amplitude of pulsatile velocity ($a$) scaled with the flow velocity ($U_0$). After the scaling analysis, the dimensional governing equations, Eqs. \ref{eq:1},\ref{eq:2}, and \ref{eq:3}, are transformed to the their dimensionless form as Eqs. \ref{eq:6},\ref{eq:7}, and \ref{eq:8}, respectively.
\begin{equation}\label{eq:6}
	\widetilde{\nabla} \cdot \widetilde{\mathbf{U}}=0
\end{equation}
\begin{equation}\label{eq:7}
	\frac{\partial \widetilde{\mathbf{U}}}{\partial \widetilde{t}}+\widetilde{\nabla}\cdot (\widetilde{\mathbf{U}}\widetilde{\mathbf{U}})=-\widetilde{\nabla} \widetilde{p}+\frac{1}{Re}\widetilde{\nabla} \widetilde{\boldsymbol{\tau}}
\end{equation}
\begin{equation}\label{eq:8}
	\widetilde{\bm{\tau}}+De~\overset{\nabla}{\widetilde{\boldsymbol{\tau}}}=2 \Bigg(\widetilde{\mathbf{D}}+\beta De~\overset{\nabla}{\widetilde{\mathbf{D}}}\Bigg)
\end{equation}	

\section{Computational details}
In the present study, we have used an open-source toolbox `$RheoTool$' \cite{rheoTool} for the numerical computations of viscoelastic fluid flows in the presence of inlet velocity pulsation. The governing conservation equations, Eqs. \ref{eq:1} and \ref{eq:2}, along with the constitutive relation, Eq. \ref{eq:3}, are solved by fully implicit finite-volume method (FVM) based on a time-marching pressure-correction algorithm \cite{oliveira1998,alves2003a} in $\text{OpenFOAM\textsuperscript{\textregistered}}$ framework. We have adopted the Semi-Implicit Method for Pressure-Linked Equations-Consistent (SIMPLEC) algorithm \cite{pimenta2017,van1984} for pressure-velocity coupling. To increase the stability of the solver, a new momentum-stress coupling algorithm (log-conformation tensor approach) is implemented in the flow solver \cite{pimenta2017}. The detailed explanation for the log-conformation tensor approach is presented in the Suplementary material.

\subsection{Boundary conditions:}
In order to solve the aforementioned dimensionless governing equation numerically, we impose a time varying pulsating axial flow velocity variation at the inlet boundary and both pressure and polymeric extra-stress components are set to zero-gradient. At the outlet, pressure is set to atmospheric whereas, all other variables are assigned as zero-gradient. No-slip and no-penetration stationary wall boundary condition is imposed to both the top and bottom walls. At these walls, pressure is set to zero normal gradient whereas, polymeric extra-stress components are linearly extrapolated.

\subsection{Grid independence analysis}\label{Grid}
{In order to ensure that the numerical results are independent of the size of the user-generated grids, a grid sensitivity test has been demonstrated by generating three different mesh structures, namely M1, M2, and M3, as given in Table \ref{Table:GCI}. In this study, the $blockMesh$ utility tool is employed to generate mesh for the simulations, which is available in `OpenFOAM\textsuperscript{\textregistered}. For distributing mapped cells in the domain, the computational domain has been split into two zones, as shown in Fig. \ref{fig:2}(a). The subplot (b) of Fig. \ref{fig:2} shows the non-uniform grid used to fully resolve the flow field in the vicinity of the channel wall. In Table \ref{Table:GCI}, $n_x$ and $n_y$ represent the number of cells in $x$ and $y$ directions, respectively in each zone and $N_c$ represents the total number of cells in the computational domain. Figure \ref{fig:2}(c) shows the axial velocity profiles along the transverse direction at $\widetilde x=40$ for three different meshes at $De=10$. From Table \ref{Table:GCI} and Fig. \ref{fig:2}(c), the relative variation in peak axial velocity is seen to be less than $1\%$ between the meshes M2 and M3. Hence, the mesh M2 is seemed as grid-independent and taken for the present computations. To cross verify the grid sensitivity test, the ``Grid Convergence Index'' (GCI) method \cite{celik2008,rath2019,rath2021} is also employed in this study. The details about the GCI algorithm and its results are provided in the Supplementary material.

\subsection{Model validation}
{In order to check the accuracy and reliability of the present numerical model, the present numerical result is validated with the existing analytical solution of Na and Yoo \cite{na1991} for a uniform flow of viscoelastic fluid in a parallel plate channel. Figure \ref{fig:3} represents the variation of axial velocity, axial normal stress, and shear stress profiles along the transverse direction at $\widetilde{x}=40$ for $Re=1$, $\beta=0.01$, and $De=0.1$. The present numerical result shows a close agreement with Na and Yoo \cite{na1991} within $1\%$ deviation. Hence, it can be believed that the numerical model implemented here would be able to predict much accurate solution for the present study.}

\section{Results and discussion}
{In this study, the time-averaged velocity and polymeric stress component profiles along the transverse and axial directions are presented to elucidate the effects of $De$, $\beta$, $Wo$, and $A$ on the flow characteristics associated with the interplay of pulsatile flow and viscoelastic nature of the fluid. However, before going to discuss the results, it is worthwhile to understand the nature of the pulsation parameters, i.e., $Wo$ and $A$. Thus, in Fig. \ref{fig:4}(a,b), the pulsation of axial velocity is denoted at a probe location of $\widetilde{x}=40,\widetilde{y}=+0.5$ in the flow domain to explain the effect of $Wo$ and $A$, respectively for $\beta=0.1$ and $De=1$. In Fig. \ref{fig:4}(a), pulsation shows for different frequencies having same amplitude of $A=0.1$ for four seconds which denotes two complete cycles of $Wo=3.545$ or $\omega=4\pi$ , four complete cycles of $Wo=2.507$ or $\omega=2\pi$, and eight complete cycles of $Wo=1.772$ or $\omega=\pi$. Similarly, in Fig. \ref{fig:4}(b), pulsation shows for different amplitudes having the same frequency of $Wo=2.507$ or $\omega=2\pi$ for two seconds which denotes two complete cycles. From Fig.\ref{fig:4}, it is clear that with an increase in $Wo$, the frequency of the axial velocity pulsation increases and a higher $A$ enhances the magnitude of $\widetilde U_x$. In the forthcoming sections, we will discuss the effects of changing fluid properties and pulsation parameters on velocity and stress distributions in the channel.}
	
\subsection{Effect of fluid elasticity}
{Unlike the Newtonian fluid where the stress is a linear function of the rate of strain, the viscoleastic fluid exhibit a non-linear time dependent relation between the fluid stress and the rate of strain. The viscoelastic fluid model constitute of two-time scales, namely relaxation time $\lambda$ and retardation time $\lambda_r$, which control the elastic behaviour of the fluid. In the presence of flow pulsation, it is expected that these fluid properties may significantly alter the flow dynamics. In Fig. \ref{fig:5}(a,b), we have highlighted the axial velocity variation for varying $De$ (non-dimensional representation of relaxation time $\lambda$) at $\widetilde x=40$ for two different $\beta$ values. From Fig. \ref{fig:5}(a), we observe that for a constant $\beta$ of 0.1, the centerline velocity increases with an increase in $De$ due to the increased elasticity of the fluid. Further, keeping the $De=50$, the variation of axial velocity with changing $\beta$ is highlighted in Fig. \ref{fig:5}(b). This figure shows that for higher $\beta$, we obtain a low centerline velocity, as compared to the low $\beta$ value. This is because as the $\beta$ increases, the elastic behaviour of the fluid diminishes, and at $\beta=1$, the viscoelastic fluid becomes Newtonian fluid. The polymeric shear stress $\widetilde \tau_{xy}$ variation along the $\widetilde y$ direction at $\widetilde{x}=40$ for changing $De$ and $\beta$ is illustrated in Fig. \ref{fig:5}(c,d) implies that for a lower $\beta$ the magnitude of $\widetilde \tau_{xy}$ at the channel wall is higher and zero at the centerline. The value of $|\widetilde \tau_{xy}|$ reduces with increasing $\beta$ and becomes zero for $\beta=1$, as the viscoelastic fluid behaves as a Newtonian fluid. The smooth variation of $\widetilde \tau_{xy}$ along $\widetilde y$ is altered with increased fluid elasticity which is evident from the comparison of $\widetilde \tau_{xy}$ vs $\widetilde y$ plots for $De=1$ and $De=50$. With increasing $De$, the fluid relaxation time increases, and with reducing $\beta$, the polymeric component of the viscosity increases. Thus, the elastic behavior of the viscoelastic fluid dominates over the viscous behavior near the walls at higher $De$ and lower $\beta$. This makes the flow characteristics more complex and as a result, the polymeric stress profiles become more bump near the walls.}

\subsection{Effect of amplitude of pulsation}
{In the pulsatile flow of viscoelastic fluid, the amplitude of pulsation ($A$) is a major flow governing parameter, which can significantly alter the flow dynamics. Hence, in Fig. \ref{fig:6}, we have presented the variation of velocity and stress components for changing $A$. It is evident from Fig. \ref{fig:6}(a,b) that with the increase in pulsation amplitude, the centerline velocity decreases because of the interplay of continuous fluctuation in axial velocity and the elastic nature of the viscoelastic fluid. It is interesting to note that for $\beta=0.9$, which denotes less elastic behaviour of fluid than $\beta=0.1$, application of equal amplitude of axial velocity fluctuation gives minimal deviations in the velocity distributions as evident from Fig. \ref{fig:8}(a). Thus, it is clear that for high $De$ or low $\beta$ the alterations in fluctuation amplitude affects the velocity distribution considerably. The decrease in the centerline velocity for higher $A$ values is also evident from the variation of $\widetilde U_x$ along the axial direction, as illustrated in Fig. \ref{fig:6}(b). Figure \ref{fig:6}(c-e) presents the variations in the time-averaged polymeric stress components at $\widetilde{x}=40$ for changing pulsation amplitude. From this figure, we observe that the magnitude of polymeric stresses is high at the near-wall region, which is generated because of the large velocity gradients at that region. We also observe that the magnitude of $\widetilde \tau_{xx}$ and $\widetilde \tau_{xy}$ are of $O(1)$, whereas the magnitude of $\widetilde \tau_{yy}$ is of the order $\sim O(10^{-4})$ at the channel walls.}
	
\subsection{Effect of pulsation frequency or effect of Womersley number}
	
{In this section, we highlight the effect of pulsation frequency, the dimensionless representation of which is the Womersley number $Wo$, on the overall flow dynamics. The alterations in the velocity and stress components with varying $Wo$ is illustrated in Fig. \ref{fig:7}. In the subplot (a) of Fig. \ref{fig:7}, we represent the axial velocity profiles along the transverse direction at $\widetilde{x}=40$. The centerline axial velocity shows a maximum value for $Wo=0$ or uniform flow condition, and it decreases marginally with the increase in $Wo$. The reduction in centerline axial velocity can also be seen in Fig. \ref{fig:7}(b), which highlights the variation of $\widetilde U_x$ as a function of $Wo$ along the axial direction. With an increase in $Wo$, the frequency of pulsation increases, and the interplay of non-zero fluid elasticity with the frequency of pulsation manifests retardation in the flow velocity. This can be further clarified from Fig. \ref{fig:8}(b), which shows the velocity distribution for exact variations in pulsation frequency at a higher $\beta=0.9$. The comparison between the two figures, i.e. Fig. \ref{fig:7}(a) and Fig. \ref{fig:8}(b) shows that for high $\beta$ or low $De$ the alterations in pulsation frequency does not affect the velocity distribution considerably as the elastic behaviour of the fluid diminishes at higher $\beta$ or lower $De$ values. From Fig. \ref{fig:7}(b), it can be observed that at the entry of the channel, the boundary layer develops, and the flow becomes fully developed after the developing region. Thus, the axial velocity increases along the axial direction up to the entry length and beyond which, it becomes constant. This phenomenon is instantaneous due to the pulsatile nature of the flow. The time-averaged polymeric stress components at $\widetilde{x}=40$ for changing pulsation frequency can be seen in Fig. \ref{fig:7}(c-e). From this figure, it is evident that the near-wall region of high-velocity gradients is the regions for high polymeric stresses and the magnitude of polymeric stresses increases with increasing pulsation frequency or $Wo$. We also observe that the magnitude of $\widetilde \tau_{xx}$ and $\widetilde \tau_{xy}$ are of $O(1)$, whereas the magnitude of $\widetilde \tau_{yy}$ is of the order $\sim O(10^{-7})$ at the channel walls. Detailed discussion on the effect of both pulsation frequency and pulsation amplitude at high retardation ratio and high Deborah number can be found form \hyperlink{A}{Appendix A} and \hyperlink{B}{Appendix B}, respectively.}
	
\section{CONCLUSIONS}			
{To conclude, we have analyzed the pulsatile flow of viscoelastic fluid in a parallel plate channel numerically to elucidate the combined effects of pulsation parameters and complex rheology of the fluid on the flow characteristics. The numerical solution is obtained by a fully implicit finite-volume method (FVM) based on a time-marching pressure-correction algorithm in $\text{OpenFOAM\textsuperscript{\textregistered}}$ framework. The grid independence study and the solver validations were carried out to check the accuracy of obtained numerical results. With increasing viscoelastic behaviour of the fluid, we observe a significant enhancement in the magnitude of axial velocity and polymeric shear stress at the near-wall region. The elasticity of the viscoelastic fluid can be enhanced either by increasing the Deborah number $De$ or by reducing the retardation ratio $\beta$. For a time-averaged pulsatile flow, we observe a reduction in the axial velocity with increasing pulsation amplitude or frequency. The extent of reduction in velocity and polymeric stress magnitude with changing pulsation parameters is highly dependent on the elastic properties of the viscoelastic fluid. The results of the present work may find its application in the transport of polymeric solutions, extrusion, and injection moulding of polymer melts in several process industries, as well as transport of biofluids in microfluidic devices.}
	
\section*{Supplementary Material}\label{sec:Supplementary_Material}
The supplementary material contains a detailed explanation of the log-confirmation tensor approach and the solution steps for grid independence study using the GCI method for the present problem.
	
\section*{Data availability statement}
The data that support the findings of this study are available from the corresponding author upon reasonable request.
\newpage
	
\begin{nomenclature}
	\begin{table}[ht]
		\hspace{-0.5cm}
		\setlength{\tabcolsep}{1em}
		\setlength{\extrarowheight}{0.0001em}
		\begin{tabular}{r l}
			$a$ & amplitude of pulsation, (m/s)\\
			$A$ & dimensionless amplitude\\				$\mathbf{D}$ & deformation rate tensor, (s$^{-1}$)\\
			$De$ & Deborah number\\  
			$H$ & half height of the channel, (m)\\
			$L$ & length of the channel, (m)\\
			$\mathbf{M}$ & arbitary second-order tensor\\
			$n$ & number of cells\\
			$N$ & total number of cells\\
			$p$ & fluid pressure, (Pa)\\
			$Re$ & Reynolds number\\
			$t$ & time, (s)\\
			$u,v$ & velocity components, (m/s)\\
			$\mathbf{U}$ & fluid velocity, (m/s)\\
			$Wi$ & Weisenberg number\\
			$Wo$ & Womersley number\\
			$x,y$ & space coordinates, (m)         
		\end{tabular}
	\end{table}    
	\subsection*{Greek symbols}    
	\begin{table}[ht]
		\setlength{\tabcolsep}{1em}
		\setlength{\extrarowheight}{0.0001em}
		\begin{tabular}{r l}
			$\beta$ & retardation ratio\\
			$\eta$ & total viscosity of the fluid, (Pa.s)\\
			$\eta_p$ & polymeric viscosity, (Pa.s)\\
			$\eta_s$ & solvent viscosity, (Pa.s) \\ 
			$\lambda$ & relaxation time, (s)\\
			$\lambda_r$ & retardation time, (s)\\
			$\omega$ & angular frequency, (s$^{-1}$)\\
			$\rho$ & fluid density, (kg/m$^3$)\\
			$\mathbf{\tau}$ & stress tensor, (Pa)   
		\end{tabular}
	\end{table}    
	\subsection*{Subscripts and superscripts}    
	\begin{table}[ht]
		\setlength{\tabcolsep}{1em}
		\setlength{\extrarowheight}{0.0001em}
		\begin{tabular}{r l}
			$0$ & mean or base\\  
			$max$ & maximum\\
			$p$ & polymeric\\ 
			$s$ & solvent\\ 
			$T$ & transpose\\ 
			$x,y$ & coordinate directions\\
			$\sim$ & non-dimensional quantities
		\end{tabular}
	\end{table}   
\end{nomenclature}
	
\newpage
\section*{Appendix}
\appendix
\section{Axial velocity distribution for changing pulsation parameters at high  retardation ratio}\hypertarget{A}
{In the present flow configuration, the interplay of fluid elasticity and the pulsation parameters gives rise to alterations in the velocity and stress distributions. The variations in the pulsation parameters alone for a Newtonian fluid will not show any deviation in the velocity or stress profiles in a time-averaged flow field. Here, for a viscoelastic fluid having high $\beta$ and low $De$ values, we have shown the axial velocity variation by changing the amplitude and frequency of pulsation in Fig. \ref{fig:8}. For $\beta=0.9$ and $De=0.1$, the amplitude of pulsation is varied from $A=0 \to 0.5$ and we observe small deviations in the centerline velocity only (refer to Fig. \ref{fig:8}(a)), which is expected because of the non-zero $De$. We have also altered the pulsation frequency, i.e. $Wo=0 \to 3.545$ keeping constant fluid elastic properties and the deviations in the velocity profile is found to be negligible (refer to Fig. \ref{fig:8}(b)). From the above discussion, it is clear that the existence of fluid elasticity enhances the effect of pulsation parameters in a pulsatile flow of viscoelastic fluid.}

\section{Axial velocity distribution for changing pulsation parameters at high  Deborah number}\hypertarget{B}
{ The viscoelastic fluid flow is highly dependent on the elastic behaviour of the fluid which are non-dimensionally denoted by Deborah number $De$ and retardation ratio $\beta$. The effect of changing $\beta$ on the axial velocity distribution is previously discussed. Here, we focus on highlighting the effect of high $De$ on the axial velocity distribution for changing $A$ and $Wo$. We have shown the axial velocity variation by changing the amplitude and frequency of pulsation in subplots ($a$) and ($b$) of Fig. \ref{fig:9}, respectively. For a higher Deborah number i.e. $De=100$, we observe a reduction in the centerline velocity with increasing the value of $A$ (refer to Fig. \ref{fig:9}(a)). However it is observed that the axial velocity is enhanced at the near wall regions to satisfy the continuity. With alterations in the pulsation frequency, i.e. $Wo=0 \to 3.545$ for a higher $De$, it is found that the higher $Wo$ is responsible for increasing the axial velocity at the near wall region and decreases the centerline velocity as seen from Fig. \ref{fig:9}(b). It is clear from the above discussion that the effect of pulsation parameters in a pulsatile flow of viscoelastic fluid is enhanced due to the increased elastic property of the working fluid.}
	
	
	\clearpage
	\newpage
	\bibliography{Manuscript}
	\bibliographystyle{asmems4}
	
	
	\newpage
	\listoffigures
	\newpage
	\listoftables
	%
	%
	\newpage
	\section*{FIGURES}
	\label{sect_figure}
	\begin{figure}[h] 
		\centering
		\includegraphics[width=0.9\textwidth]{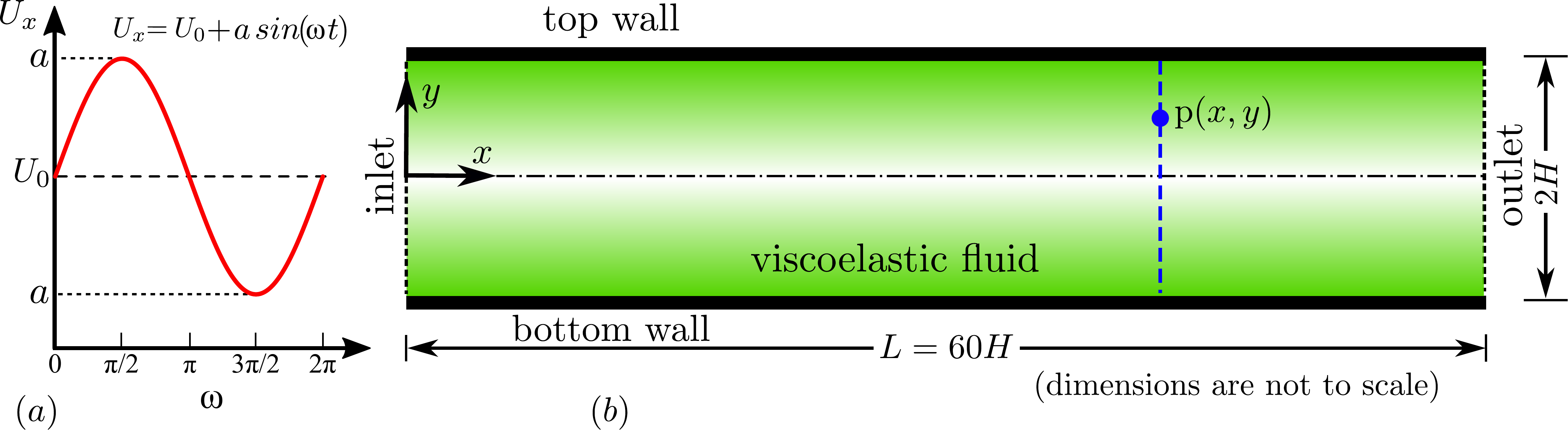}
		\caption{(Color Online) Schematic of the problem; (a) axial pulsation representation of the flow for a particular case of unity frequency; (b) computational domain  of the present flow configuration where the point $p(x,y)$ denots the probe location}
		\label{fig:1}
	\end{figure}
	\newpage
	\begin{figure}[h]
		\centering
		\includegraphics[width=0.85\textwidth]{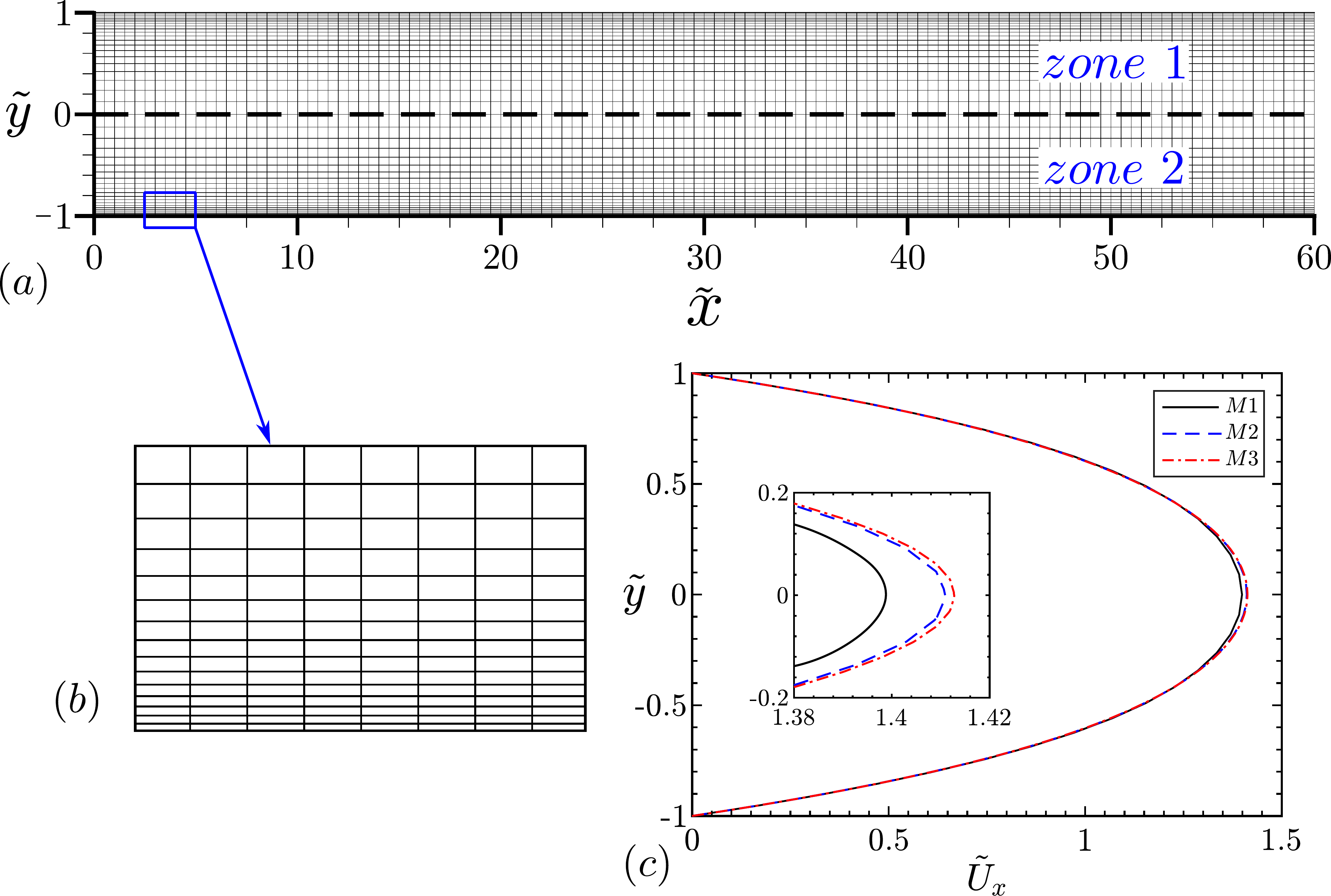}
		\caption{(Color Online) (a) Grid resolution of the computational domain which is divided into two zones, (b) magnified view of the near wall grid region, (c) variation of axial velocity with changing mesh resolution along the transverse direction at $\widetilde x=40$ for $De=10$, $\beta=0.1$, $Wo=1.772$, and $A=0.5$}
		\label{fig:2}
	\end{figure}
	\newpage
	\begin{figure}[ht]
		\centering
		\includegraphics[width=0.5\textwidth]{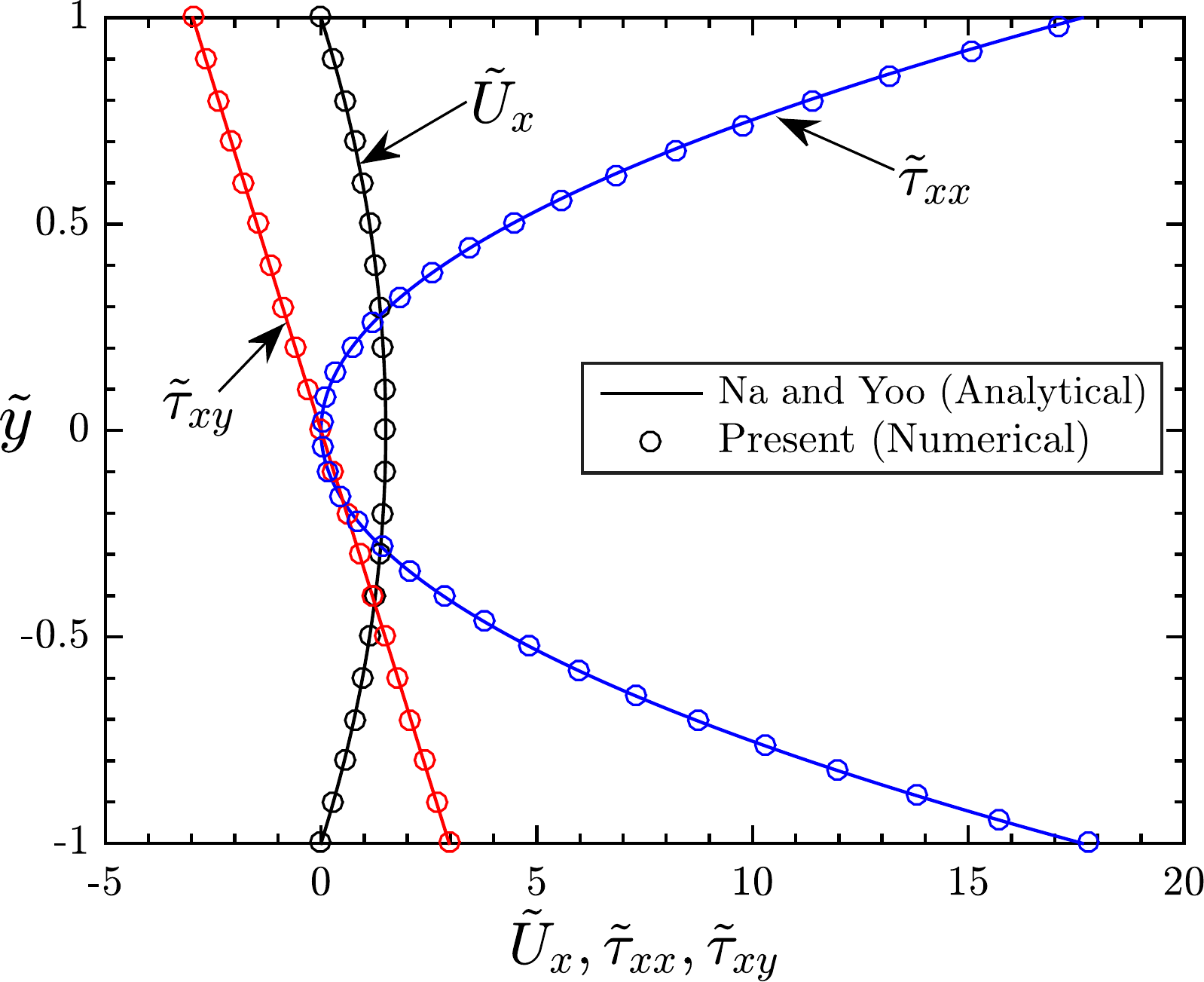}
		\caption{(Color Online) Validation with the analytical work of Na and Yoo \cite{na1991} for a uniform flow having $Re=1$, $\beta=0.01$, and $De=0.1$, representing axial velocity and polymeric stress component profiles along the transverse direction at $\widetilde{x}=40$}
		\label{fig:3}
	\end{figure}
	\newpage
	\begin{figure}[h]
		\centering
		\includegraphics[width=0.85\textwidth]{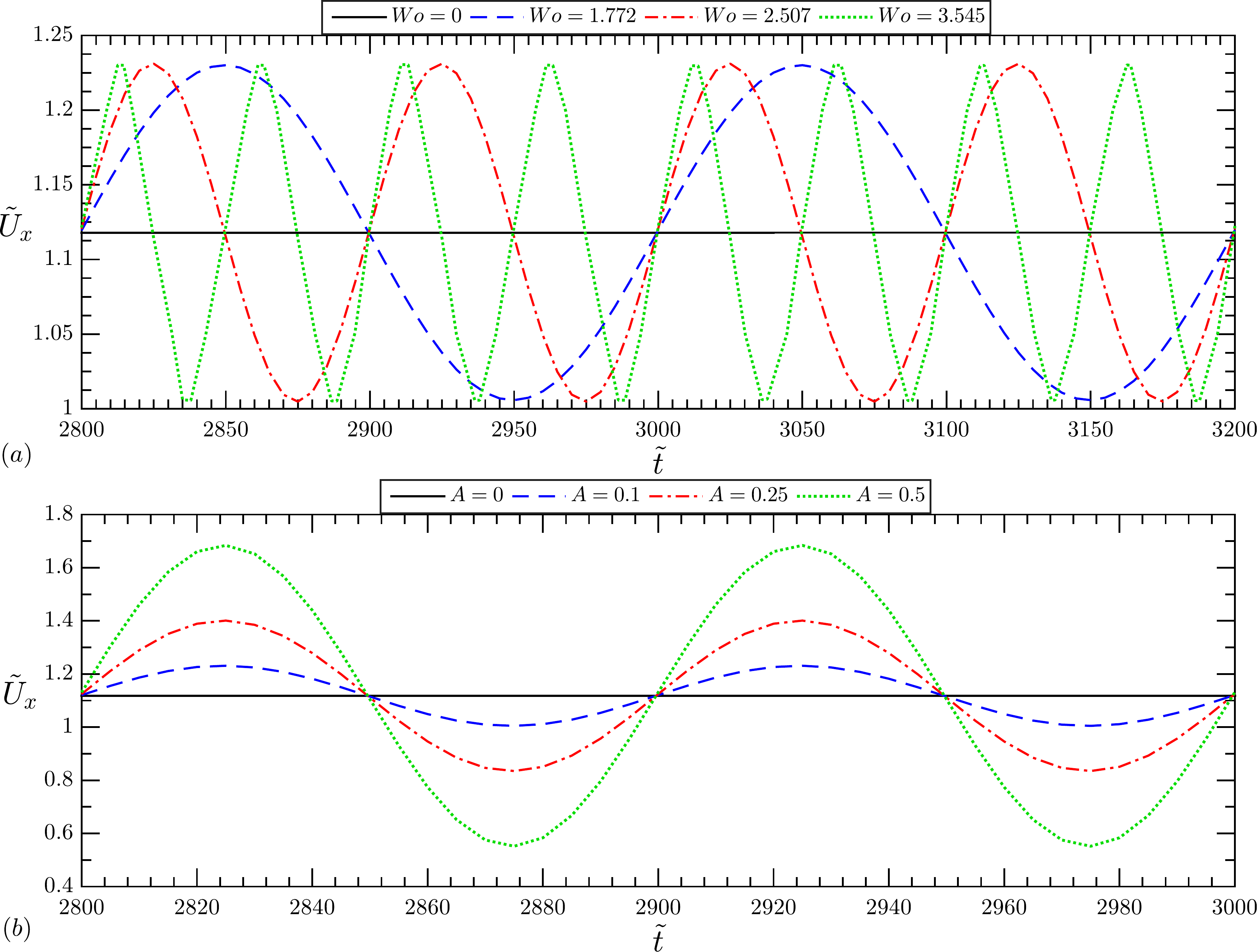}
		\caption{(Color Online) Variation of axial velocity at a probe location ($\widetilde{x}=40,\widetilde{y}=+0.5$) in the channel with time for (a) varying $Wo$ at $A=0.1$, (b) varying $A$ at $Wo=2.507$; other simulation parameters are $\beta=0.1$ and $De=1$}
		\label{fig:4}
	\end{figure}
	\newpage
	\begin{figure}[h]
		\centering
		\includegraphics[width=0.85\textwidth]{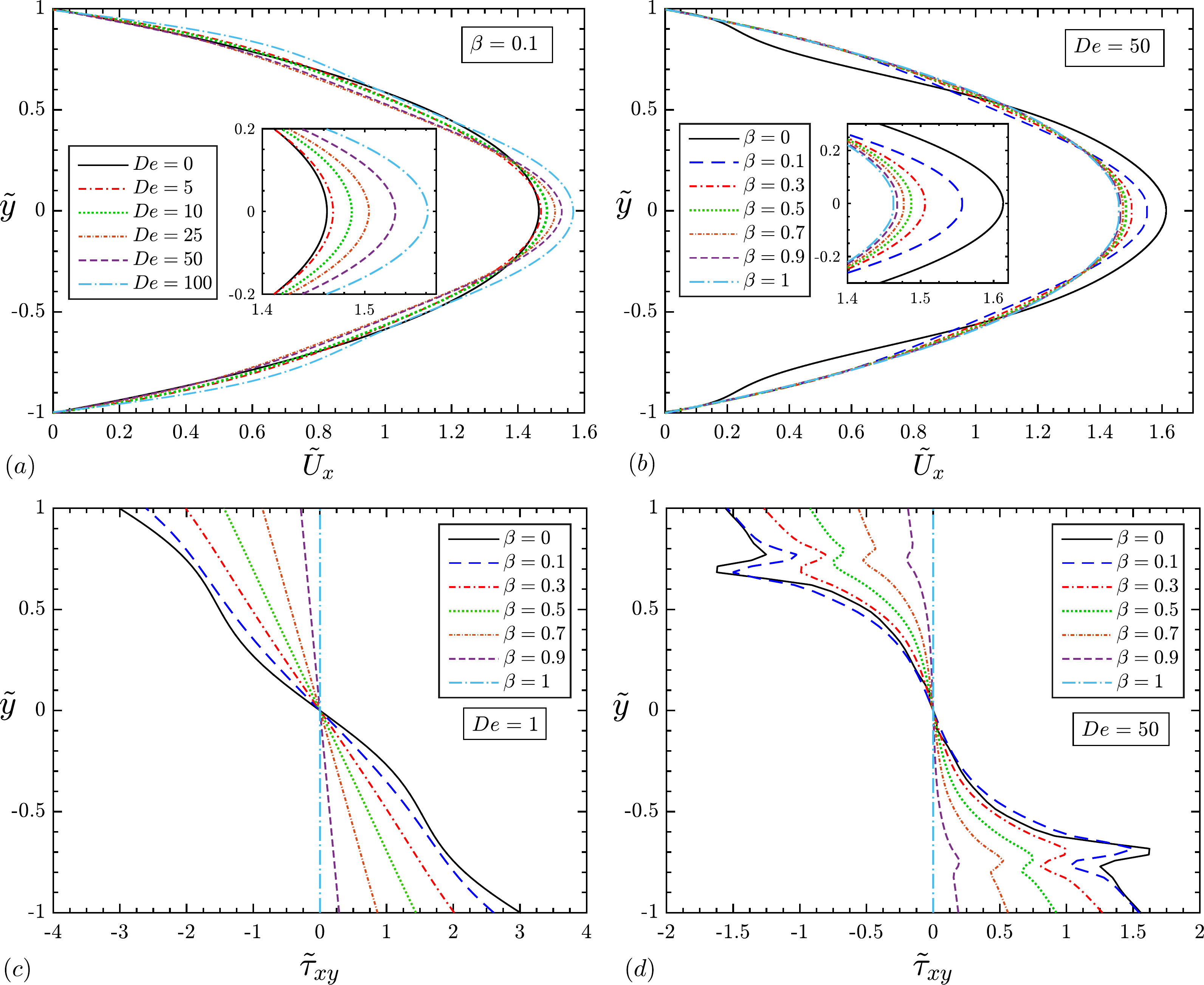}
		\caption{(Color Online) (a) Variation of the axial velocity along the transverse for varying $De$ at $\widetilde{x}=40$ for $Wo=2.507$, $\beta=0.1$, and $A=0.1$; (b) Variation of the axial velocity along the transverse for varying $\beta$ at $\widetilde{x}=40$ for $Wo=2.507$, $De=50$, and $A=0.1$; (c) $\widetilde \tau_{xy}$ vs $\widetilde y$ for varying $\beta$ at $\widetilde{x}=40$, $A=0.25$, $De=1$, and $Wo=2.507$; (d) $\widetilde \tau_{xy}$ vs $\widetilde y$ for varying $\beta$ at $\widetilde{x}=40$, $A=0.25$, $De=50$, and $Wo=2.507$ }
		\label{fig:5}
	\end{figure}
	\newpage
	\begin{figure}[h]
		\centering
		\includegraphics[width=0.85\textwidth]{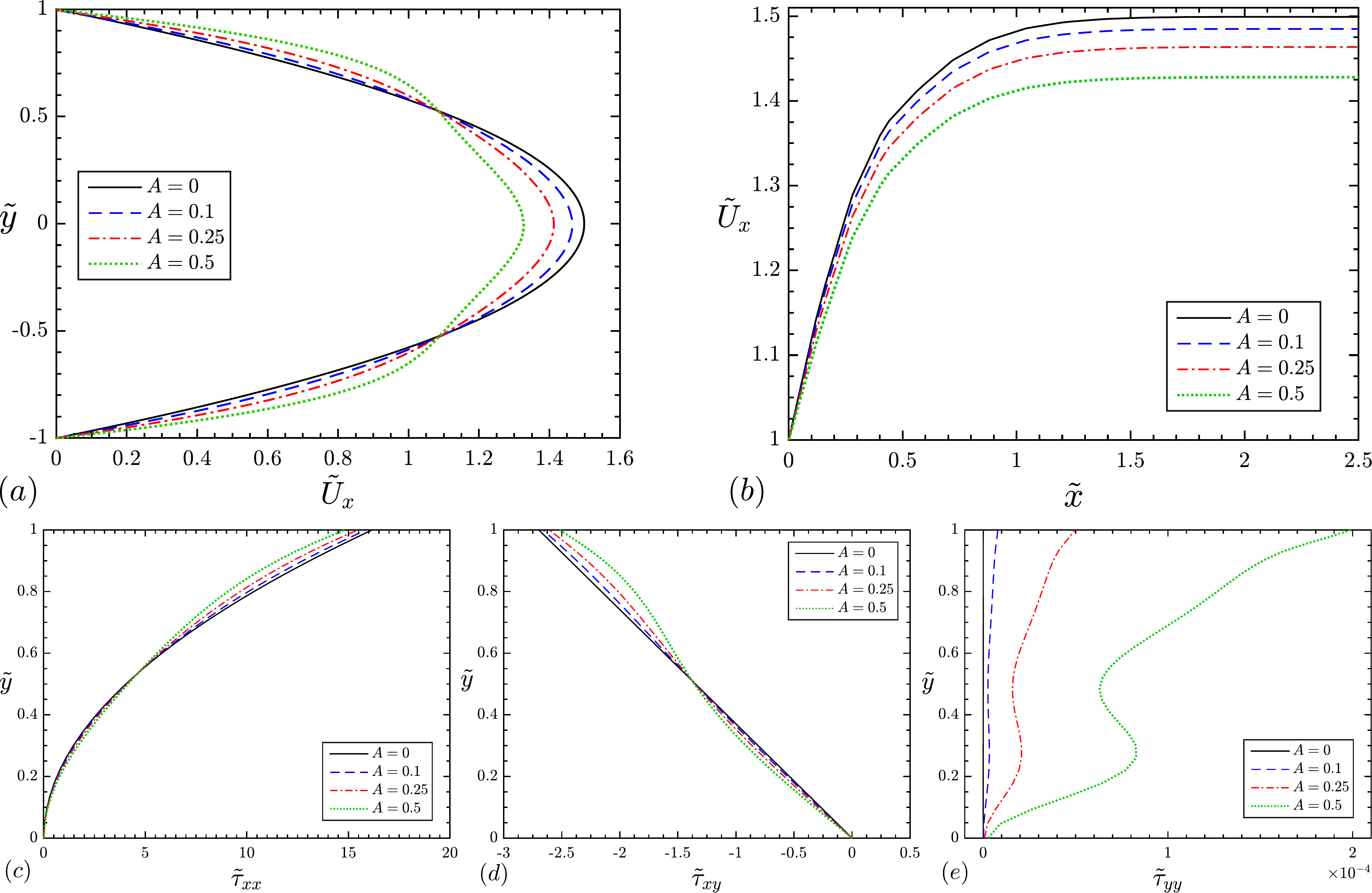}
		\caption{(Color Online) Variation of velocity and stress componenets with changing $A$: (a) $\widetilde U_x$ vs $\widetilde y$ at $\widetilde{x}=40$, $De=0.1$, $\beta=0.1$, and $Wo=2.507$; (b) $\widetilde U_x$ vs $\widetilde x$ at $\widetilde{y}=0$, $De=0.1$, $\beta=0.1$, and $Wo=2.507$; (c) variation of axial normal stress along the transverse direction, i.e., $\widetilde \tau_{xx}$ vs $\widetilde y$; (d) variation of shear stress along the transverse direction, i.e., $\widetilde \tau_{xy}$ vs $\widetilde y$; (e) variation of transverse normal stress along the transverse direction, i.e., $\widetilde \tau_{yy}$ vs $\widetilde y$ at $\widetilde{x}=40$, $De=0.1$, $\beta=0.1$, and $Wo=2.507$}
		\label{fig:6} 
	\end{figure}
	\newpage
	\begin{figure}[ht]
		\centering
		\includegraphics[width=0.85\textwidth]{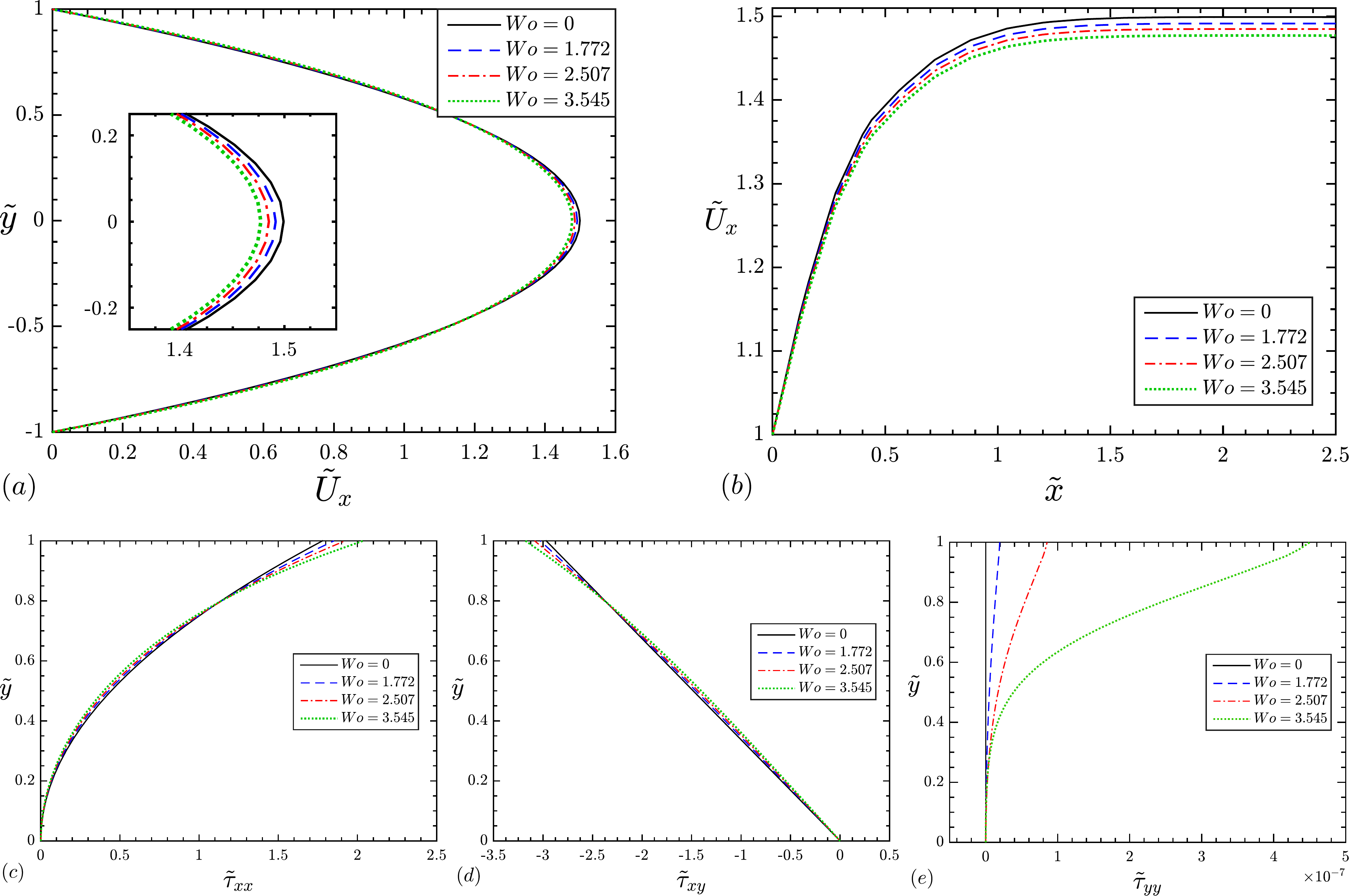}
		\caption{(Color Online) Variation of velocity and stress componenets with changing $Wo$: (a) $\widetilde U_x$ vs $\widetilde y$ at $\widetilde{x}=40$, $De=0.1$, $\beta=0.1$, and $A=0.1$; (b) $\widetilde U_x$ vs $\widetilde x$ at $\widetilde{y}=0$, $De=0.1$, $\beta=0.1$, and $A=0.1$; (c) variation of axial normal stress along the transverse direction, i.e., $\widetilde \tau_{xx}$ vs $\widetilde y$; (d) variation of shear stress along the transverse direction, i.e., $\widetilde \tau_{xy}$ vs $\widetilde y$; (e) variation of transverse normal stress along the transverse direction, i.e., $\widetilde \tau_{yy}$ vs $\widetilde y$ at $\widetilde{x}=40$, $De=0.1$, $\beta=0.1$, and $A=0.1$}
		\label{fig:7}
	\end{figure}
	\newpage
	\begin{figure}[h] 
		\centering
		\includegraphics[width=0.85\textwidth]{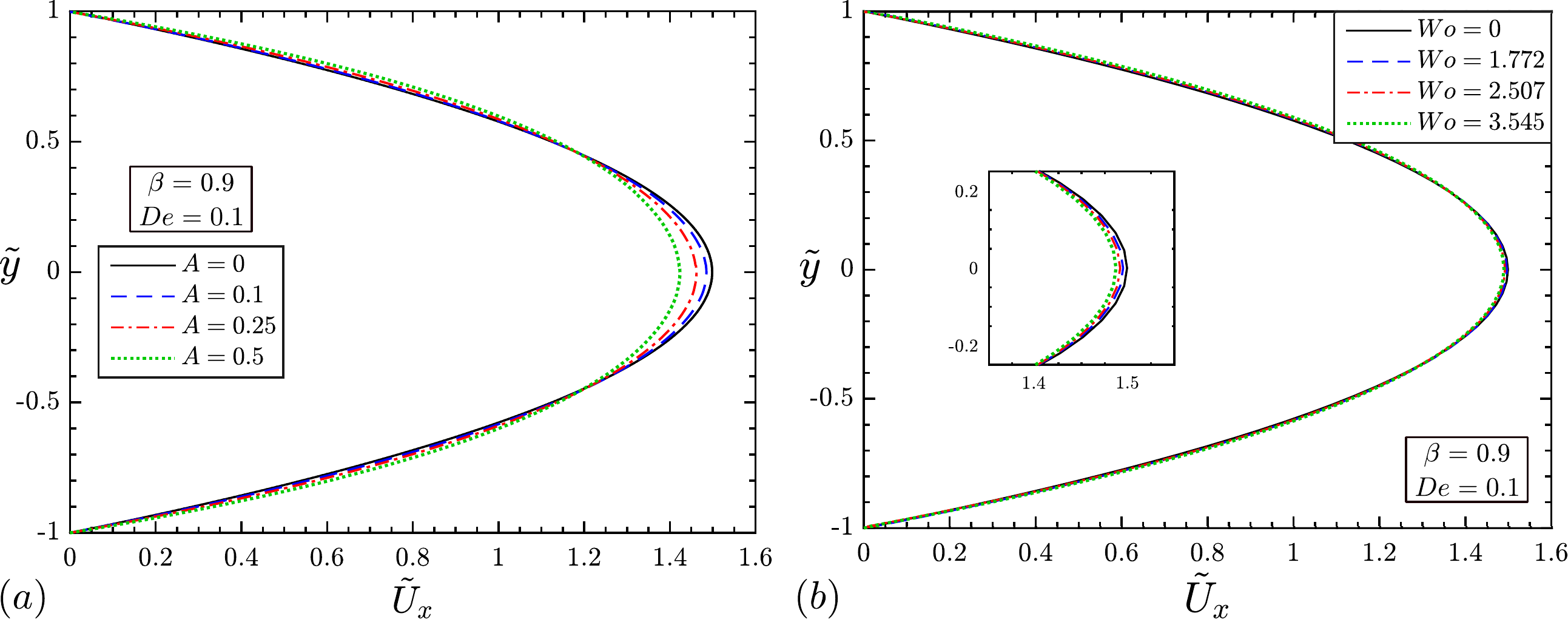}
		\caption{(Color Online) (a) Variation of axial velocity with changing $A$, i.e., $\widetilde U_x$ vs $\widetilde y$ at $\widetilde{x}=40$, $De=0.1$, $\beta=0.9$, and $Wo=2.507$; (b)  Variation of axial velocity with changing $Wo$, i.e., $\widetilde U_x$ vs $\widetilde y$ at $\widetilde{x}=40$, $De=0.1$, $\beta=0.9$, and $A=0.1$}
		\label{fig:8}
	\end{figure}
    \newpage
    	\begin{figure}[h] 
    	\centering
    	\includegraphics[width=0.85\textwidth]{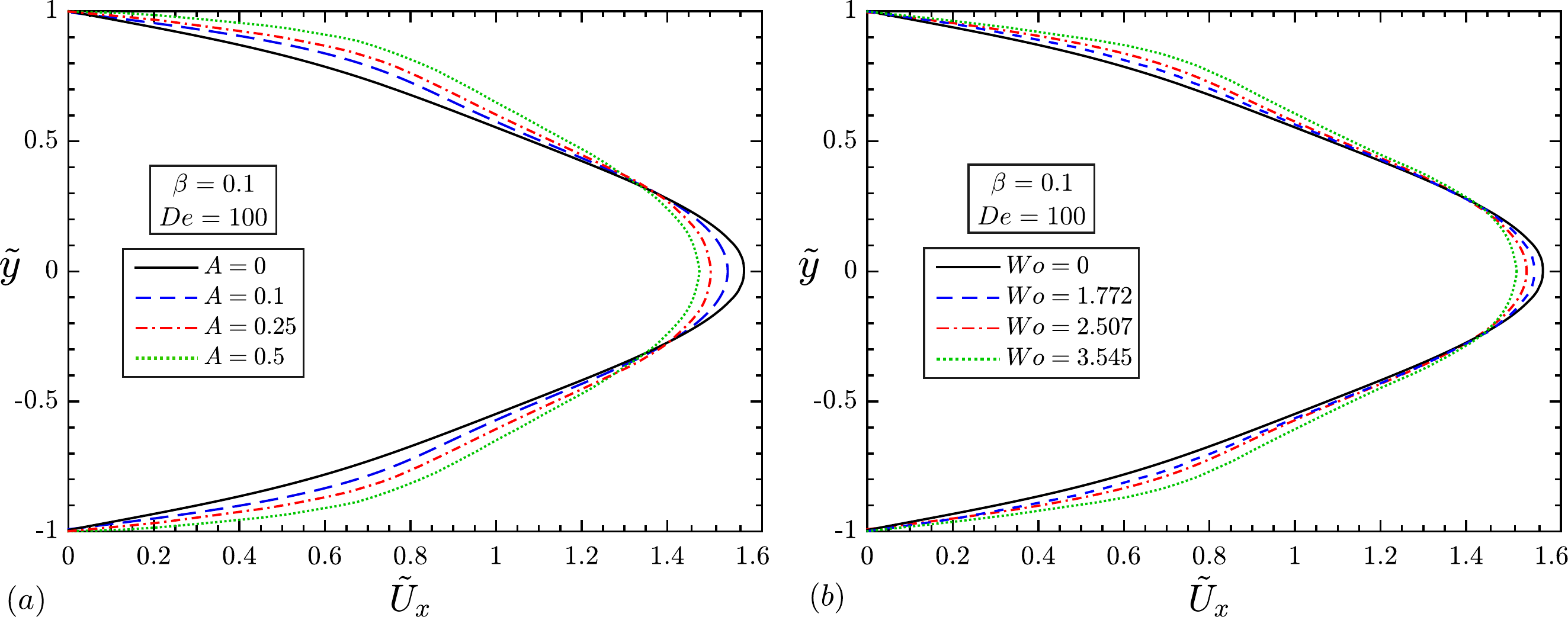}
    	\caption{(Color Online) (a) Variation of axial velocity with changing $A$, i.e., $\widetilde U_x$ vs $\widetilde y$ at $\widetilde{x}=40$, $De=100$, $\beta=0.1$, and $Wo=2.507$; (b)  Variation of axial velocity with changing $Wo$, i.e., $\widetilde U_x$ vs $\widetilde y$ at $\widetilde{x}=40$, $De=100$, $\beta=0.1$, and $A=0.1$}
    	\label{fig:9}
    \end{figure}
	%
	\newpage
	\section*{TABLES}
	%
	\begin{table}[h]
		\caption{Range of values of the non-dimensional quantities used in the numerical solution}
		\centering
		\setlength{\tabcolsep}{1.5em}
		\setlength{\extrarowheight}{1.5pt}
		\begin{tabular}{llll}
			\hline \hline
			Symbols & Name            & Definition                                                                                                    & Range of values \\
			\hline
			$Re$      & Reynolds number   & $\frac{\rho U_0H}{\eta}$                                                                                        & $1$                 \\
			$De$      & Deborah number    & $\frac{\lambda U_0}{H}$                                                                                         & $0-100$             \\
			$\beta$   & Retardation ratio & $\frac{\lambda_r}{\lambda}=\frac{\eta_s}{\eta_s+\eta_p}$ & $0-1$               \\
			$Wo$      & Womersley number     & $H\sqrt{\frac{\omega\rho}{\eta}}$                                                                    & $0-3.545$           \\
			$A$       & Pulsation amplitude   & $\frac{a}{U_0}$                                                                                                 & $0-0.5$  \\      
			\hline \hline    
		\end{tabular}
		\label{tab:parameters}
	\end{table}
	\newpage
	
	\begin{table}[h]
		\caption{Characteristics of blockMesh and grid independence test showing in terms of maximum peak velocity (Note: mesh in z-direction is having unity depth i.e., $n_z=1$ and grading value= 1)}\label{Table:GCI}
		\begin{tabular}{lccccccc}
			\hline\hline
			{Mesh} & {~~~~~~~~~~Zone 1} &{}     & {~~~~~~~~~~Zone 2} &{}     & {$N_c$} & {$\widetilde {U}_{x,max}$} & {$\%~error$} \\ \cline{2-3}\cline{4-5}
			& $n_x\times n_y$ & Grading & $n_x\times n_y$ & Grading &                        &                                          &                          \\ \hline
			M1                    & $50\times 20$   & $[10~~0.5]$     & $50\times 20$   & $[10~~2]$   & 2000                   & 1.3956                                   & -                        \\
			M2                    & $100\times 40$   & $[10~~0.5]$     & $100\times 40$   & $[10~~2]$   & 8000                   & 1.4111                                   & 1.11                     \\
			M3                    & $150\times 60$  & $[10~~0.5]$     & $150\times 60$  & $[10~~2]$   & 18000                  & 1.4148                                   & 0.26                     \\ \hline\hline
		\end{tabular}
	\end{table}

\end{document}